\documentclass[runningheads]{llncs}
\usepackage[T1]{fontenc}
\usepackage{graphicx}
\newif\ifblindit
\blinditfalse
\newcommand{\blindit}[1]{%
\ifblindit
  (blinded for review)\relax
\else
  #1\relax
\fi
}
\begin{document}
\title{Efficacy of a Computer Tutor that Models Expert Human Tutors}
%
%

\author{Andrew M. Olney\inst{1}\orcidID{0000-0003-4204-6667} \and
Sidney K. D'Mello\inst{2}\orcidID{0000-0003-0347-2807} \and
Natalie Person\inst{3} \and
Whitney Cade\inst{4}\orcidID{0000-0002-6012-7789} \and
Patrick Hays\inst{1} \and
Claire W. Dempsey\inst{1} \and
Blair Lehman\inst{5}\orcidID{0000-0002-4091-0688} \and
Betsy Williams\inst{3}\orcidID{0009-0004-9425-8447} \and
Art Graesser\inst{1}\orcidID{0000-0003-0345-6866}
}
%
\authorrunning{\blindit{A. Olney et al.}}
\institute{University of Memphis, Memphis TN 38152, USA\\
\email{\{aolney,dphays,mcwllams,graesser\}@memphis.edu}\\
\and
University of Colorado Boulder, Boulder CO, 80309, USA\\
\email{sidney.dmello@colorado.edu}\\
\and
Rhodes College, Memphis, TN 38112, USA\\
\email{\{person,sandersb\}@rhodes.edu}\\
\and
American Institutes for Research, Arlington VA 22202, USA\\
\email{wcade@air.org}\\
\and
Brighter Research, Thousand Oaks, CA 91362, USA
\email{blehmann@brighter-research.com}
}
\maketitle              
\begin{abstract}
Tutoring is highly effective for promoting learning.
However, the contribution of expertise to tutoring effectiveness is unclear and continues to be debated. 
We conducted a 9-week learning efficacy study of an intelligent tutoring system (ITS) for biology modeled on expert human tutors with two control conditions: human tutors who were experts in the domain but not in tutoring and a no-tutoring condition.
All conditions were supplemental to classroom instruction, and students took learning tests immediately before and after tutoring sessions as well as delayed tests 1-2 weeks later.
Analysis using logistic mixed-effects modeling indicates significant positive effects on the immediate post-test for the ITS ($d =.71$) and human tutors ($d =.66$) which are in the 99th percentile of meta-analytic effects, as well as significant positive effects on the delayed post-test for the ITS ($d =.36$) and human tutors ($d =.39$).
We discuss implications for the role of expertise in tutoring and the design of future studies.

\keywords{intelligent tutoring systems \and expert tutor \and dialogue \and animated pedagogical agent \and biology}
\end{abstract}
%
%
\section{Introduction}\label{sec:intro}

Meta-analyses of decades of research support the effectiveness of human tutoring for promoting learning \cite{Hartley1977,Cohen1982,Mathes1994,Ritter2009,VanLehn2011}.
The median effect size (ES) across these meta-analyses is .4 over conventional instruction, which is equivalent to an improvement from the 50th percentile to the 66th percentile.
While these meta-analyses don't contrast the effectiveness of tutors by expertise, untrained tutors (.36 ES) are about as effective as trained tutors (.41 ES) \cite{Cohen1982}, and peer tutors (.52 ES) are about as effective as adults (.54 ES) \cite{Hartley1977}. 
Indeed across these meta-analyses, the median effect size when children are tutors (.4 ES) is comparable to the median effect size when adults are tutors (.38 ES).
Thus these meta-analyses bring into question the intuition that training and age should increase tutoring effectiveness, leading some to argue that tutor experience is not important as long as the tutor is sufficiently knowledgeable and interactive \cite{VanLehn2011}, which are both required for effective feedback and scaffolding.

Bloom's landmark 2-sigma paper \cite{Bloom1984}, which describes tutoring effects that are five times the median tutoring effect above --- 2 ES or an improvement from the 50th percentile to the 98th percentile --- has been viewed as evidence supporting the importance of tutoring expertise.
Bloom's paper summarizes the work of two of his students \cite{Anania1981,Burke1983} who found effect sizes of approximately 2 ES in their dissertations, which contrast tutoring, mastery learning (in which students can't move on until achieving criterion mastery), and conventional instruction.
The 2-sigma paper was instrumental in promoting the idea of ``super-tutors,'' tutors with effectiveness beyond what would be observed in typical settings.
However, as noted by others \cite{VanLehn2011,Kulik2016}, Bloom's presentation of the 2 ES effect obfuscates both that the tutoring condition in these studies includes mastery learning and that the mastery learning criterion for the tutoring condition was 90\% vs. the mastery condition criterion of 80\%.

Teasing apart the contribution of mastery learning to the 2-sigma effect is not trivial.
A meta-analysis of mastery learning found that changing the criterion from 80\% to 90\% has virtually no effect on learning \cite{Kulik1990}.
If we assume that mastery learning and tutoring are making additive contributions to learning (i.e. no interaction), then we can simply subtract the effect size of mastery learning to get the tutoring effect.
This calculation, using the actual average effect sizes in the 2-sigma studies \cite{Anania1981,Burke1983}, yields 1.91 ES for tutoring+mastery learning, 1.06 ES for mastery learning, and a .85 ES net effect of tutoring.
While .85 is a larger effect of tutoring than might be expected from meta-analyses generally, it is consistent with a meta-analysis .83 ES reported for human tutoring studies that create their own learning outcome measures when children are tutors \cite{Cohen1982}.
Additionally, both of the  2-sigma studies used undergraduate education majors as tutors, with one study further describing tutor training as lasting one week \cite{Burke1983}.
The combination of conventional effect size and lack of extensive experience of the tutors, together with the confounding of mastery learning and tutoring, suggests that the 2-sigma tutors were not ``super tutors'' after all.
Nevertheless, the 2-sigma paper has been influential in shaping the development of intelligent tutoring systems moving forward, cf. \cite{Anderson1985a}.

Intelligent tutoring systems (ITS) compare favorably to human tutors in terms of effectiveness. 
A meta-analysis of ITS effectiveness found a median .66 ES compared to conventional instruction, which is equivalent to an improvement from the 50th percentile to the 75th percentile \cite{Kulik2016}.
While a median .66 ES for ITS is larger than the median .4 ES for human tutoring described above, the meta-analysis also found that the ES depended heavily on the type of test used to measure learning.
When the test was developed by researchers, the ITS effect size was .73 ES (compared to .84 ES for child tutors with such tests \cite{Cohen1982}), but when a standardized test was used, the ITS effect size was .13 ES (compared to .27 ES for child tutors with such tests \cite{Cohen1982}).
Tests that combined researcher and standardized items had an intermediate .45 ES.
Test type was found to be the single most important predictor of ITS effect size,  such that when test type was held constant, no other study features influenced the size of the effect.
In summary, human tutors may be more effective than ITS when test type is considered, so direct comparisons of ITS and human tutors are essential for understanding their effectiveness.

Our research builds on the intuition that expertise in tutoring makes a contribution to tutoring effectiveness beyond subject matter expertise.  
This intuition is informed by our observations and analyses of expert human tutors \blindit{\cite{Olney2010b,Olney2012}}.
These expert tutors had 5+ years of tutoring experience, a teaching license, a degree in the tutored subject, and reputations in the community as effective tutors.
In contrast to novice tutors, these expert tutors were more interactive, diagnostic, and gave more discriminating feedback, all of which have been cited as theoretical reasons for the effectiveness of human tutors (for a review see \cite{VanLehn2011}).
Our goal was to build an ITS modeled on these expert human tutors, both to increase the effectiveness of the ITS and to provide further evidence on the role of expertise in tutoring.
In this paper, we present the results of a 9-week study that compared the ITS to subject matter expert tutors and a classroom control.
We use new analytical techniques to extend a previous analysis of the first 3 weeks \blindit{\cite{Olney2012a}} and interpret the results from all 9 weeks in light of meta-analyses that have since been published.
Our research questions are: (1) is learning in the ITS condition different from the classroom control and (2) is learning in the ITS condition different from subject matter expert tutors? 

\section{\blindit{Guru}: an ITS modeled on expert human tutors}

We developed an intelligent tutoring system (ITS) for high school biology called \blindit{Guru}. 
\blindit{Guru} is a dialogue-based ITS in the style of the AutoTutor ITS family \cite{Nye2014}.
Like AutoTutor, an animated tutor agent engages the student in a natural language dialogue in which the student and tutor collaboratively interact with a multimedia workspace that displays and animates images that are relevant to the conversation. 
Student responses are analyzed with natural language understanding techniques in order to provide formative feedback and tailor the dialogue to individual students' knowledge levels.
In contrast to AutoTutor, \blindit{Guru}'s pedagogical and motivational strategies are informed by in-depth observation and computational modeling of approximately 50 hours of one-on-one tutoring between 39 students and 10 expert tutors \blindit{\cite{Person2007a,Olney2010b,DMello2010b}}.
The \blindit{Guru} animated pedagogical agent is also more sophisticated than the standard AutoTutor agent and uses motion capture data to produce realistic gestures and pointing.
\blindit{Guru} was designed to cover \blindit{Tennessee} Biology I Curriculum Standards using curriculum maps provided by the state.
For example, the state standard ``Distinguish among the structure and function of the four major organic macromolecules found in living things'' is mapped to the outcome ``Describe the structure and function of lipids, carbohydrates, and proteins'' which is broken down into multiple topics including Protein Function, which has 11 concepts (e.g., proteins help cells regulate functions).
Thus \blindit{Guru}'s topic coverage is highly standardized and aligned with the state biology curriculum.
In addition to covering topics in tutorial dialogue sessions, \blindit{Guru} presents interactive tasks such as generating summaries, completing concept maps, and cloze tasks.
We next describe the structure of a tutoring session including these tasks.

\subsection{Structure of a \blindit{Guru} session}
A typical session is structured as follows: Collaborative Lecture, Summary, Concept Maps I, Scaffolding I, Concept Maps II, Scaffolding II, and Cloze Task. 
The tutoring session is structured to resemble patterns in expert human tutoring sessions \blindit{\cite{DMello2010b}}.

 \begin{figure}[htp]
 \includegraphics[width=\textwidth]{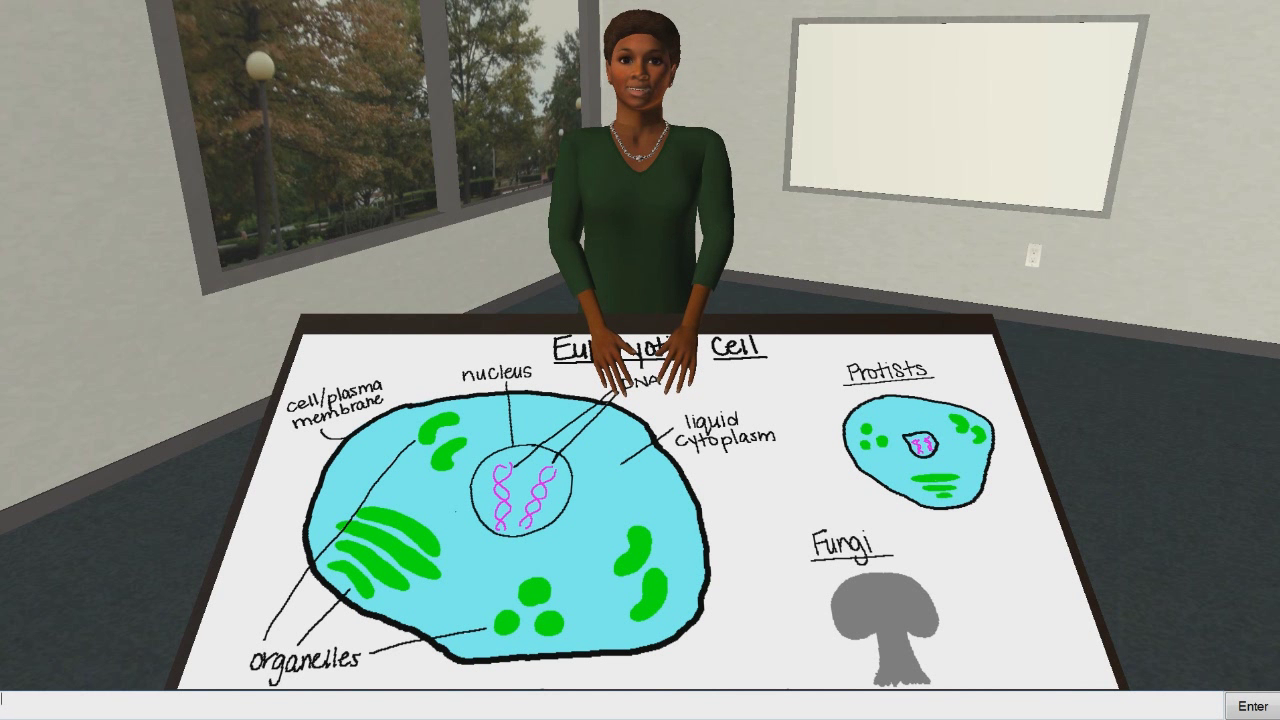}
 \caption{The \blindit{Guru} interface in Collaborative Lecture and Scaffolding modes.} \label{fig:interface}
 \end{figure}

\noindent
\textbf{Collaborative Lecture.} Collaborative Lecture is designed to cover all of the concepts for each topic and is modeled after the interactive lecture styles of expert human tutors \blindit{\cite{DMello2010}}. 
These tutoring lectures differ from typical classroom lectures in that students make frequent contributions: there is a 3:1 (Tutor:Student) turn ratio in these collaborative lectures, which represents substantially higher student participation than found in classroom lectures.
Collaborative Lecture begins with a brief preview of the topic delivered by the tutor.
When possible, the tutor relates the topic to something concrete with potential to be personally relevant to the student.  
For example, in the preview for Protein Function, the tutor says, ``Proteins do lots of different things in our bodies. In fact, most of your body is made out of proteins!''
During collaborative lecture, the tutor asks students simple concept completion questions (e.g., Enzymes are a type of what?), verification questions (e.g., Is connective tissue made up of proteins?), or comprehension gauging questions (e.g., Is this making sense so far?) to ensure the students are paying attention and are engaged with the material. 
The tutor acknowledges and responds based on the student's answers, e.g. ``Very good,''
and can also respond to student initiative statements like ``I don't understand,'' ``Can you repeat that,'' etc.
The dialogue is thus controlled using two different models, one specific to Collaborative Lecture and one that is a more general model for handling dialogue that occurs across all contexts (e.g., feedback, questions, and motivational dialogue).
The models are informed by an ensemble of speech act classifiers for determining the student's intent \blindit{\cite{Olney2009,Rasor2011}} in addition to the tutor's goals and the dialogue history.
Presentation of concepts is aided by the multimedia display as shown in Figure~\ref{fig:interface}. 
Elements are sequentially added to the multimedia display timed to the tutor's speech.  The tutor also points to elements and gestures in time with speech in order to direct the student's attention and emphasize points.

\noindent
\textbf{Student Summary.}  After the collaborative lecture, students are asked to generate a summary of what has been discussed.
Contents on the multimedia panel display are removed to make the summary a pure recall and constructive task.
The quality of the summary determines both the structure of the remainder of the session as well as which concepts will be addressed further in the Concept Map and Scaffolding phases.
If the student covers a third or less of the concepts in the summary, the session will have two rounds of Concept Maps and Scaffolding, otherwise, the session will have one round of Concept Maps and Scaffolding.
Additionally, any concepts covered in the summary are presumed to be understood by the student and so are not covered again in any session.

\noindent
\textbf{Concept Maps.} Students complete a skeleton concept map \cite{Novak1990,Novak2006} for concepts they omitted from their summaries. 
A skeleton concept map is a concept map where some nodes and/or edges have been deleted.
Students are provided with separate answer banks of nodes/edges, and when they type the correct answer for a node/edge, the corresponding entry disappears from the answer bank.
The number of skeleton concept maps for each topic is determined by the number of concepts for the topic and how many triples, e.g. $proteins \Rightarrow build \Rightarrow muscle$ are represented in the concept.
To avoid overloading the students, maps are limited to a maximum of four triples.
The skeleton concept maps are automatically generated from the text of each concept \blindit{\cite{Olney2011}}.
Because the concept maps are a recognition task, success on the concept maps is not considered as evidence that the student has learned the concept. 
The concepts are only considered covered when students can provide correct answers in recall tasks.\\  

\noindent
\textbf{Scaffolding.} After the student completes all of the concept maps, the tutoring session resumes with dialogue-based scaffolding. 
First, the multimedia display is reset to avoid providing clues to the student.
As the student demonstrates an understanding of concepts, corresponding elements are revealed on the multimedia display and remain present for the remainder of the scaffolding session.
The scaffolding dialogue covers all of the concepts that were omitted from the student-generated summaries. 
Currently, \blindit{Guru} adheres to a $Prompt \rightarrow Feedback \rightarrow Verification Question \rightarrow Feedback$ dialogue cycle to help students learn each important concept. 
The cycle for a concept is terminated as soon as the student demonstrates understanding, so the shortest possible cycle for a concept is $Prompt \rightarrow Feedback$.
Prompts and verification questions are selected by projecting their text into a vector space and then aggregating vectors across turns into an orthonormal basis \blindit{\cite{Olney2005}}.
The question that would maximize the student's assessment, if they gave the correct answer, is then selected.
The orthonormal basis is also used to check if the concept falls within the common ground of the dialogue.
If not, a Preview is generated, e.g. ``Let's talk about how our bodies use proteins,'' before the tutor asks any questions.
Student responses to questions are assessed using a combination of cosine and keyword matching with edit distance, which are
calculated against the expected answer for the question, and the assessment of the student's response is the maximum of these two calculations.
Feedback to students ranges in five levels from negative to positive.
Negative feedback is followed by an encouraging solidarity statement, e.g. ``That's OK, you'll get it.''
In addition to this tutor initiative dialogue, \blindit{Guru} can also respond to student initiative as described in collaborative lecture.
Scaffolding has its own dialogue model which can address concepts in any order, making it considerably more dynamic than collaborative lecture.\\

\noindent
\textbf {Cloze Task.}
The session concludes with an interactive Cloze task. 
Cloze tasks are activities that require students to supply missing concepts from a passage \cite{Taylor1953}. 
The passages are the ``ideal'' summaries for each topic; they include a cohesive passage that synthesizes the text from each concept. 
Students are not given an answer bank or any feedback on this test, so it can be considered a summative retrieval practice task or assessment.

\section{Method}

\subsection{Participants}
Thirty-four tenth graders from an urban high school in the U.S. volunteered to participate in the study in the fall of 2011. 
All students were enrolled in Biology I and had the same teacher for that course. 
Once a week (for nine weeks) 
students participated in the study during another class period.
It is worth noting that all students were required to pass the state-mandated end-of-course assessment for Biology I to graduate from high school.

\subsection{Design}\label{sec:design}

The study used a three-condition repeated-measures design in which each student interacted with both \blindit{Guru} (ITS) and a human tutor (Human) in addition to their regular classroom instruction (Class). The tutoring topics (for both ITS and Human) always lagged behind what the biology teacher covered in the classroom by one week.  For example, if the teacher covered Topic A (e.g. Biochemical Catalysts) one week, the ITS and Human conditions would tutor on Topic A the following week. Students were assigned to groups so that in a particular week they would receive either ITS or Human conditions, and on the following week, they would receive the opposite tutoring condition (e.g., ITS week 1, Human week 2, or vice versa). All students received classroom instruction each week.

The study design unfolded over three, 3-week cycles for a total of 9 weeks.
Each cycle addressed four topics, where two were tutored (A, B) and two were not tutored (X, Y); this enabled comparison of tutoring (A, B) vs. classroom instruction only (X, Y);
For the first two weeks of a cycle, the tutoring conditions (ITS, Human) completed immediate pre- and post-tests for both tutored and non-tutored topics (AX or BY).
On the third week of a cycle, students took a delayed post-test covering all four topics.
Table~\ref{tab:design} presents a schedule of this design for the first cycle of the study. Week 0 is considered a non-study week because researchers did not interact with participants.

\begin{table}[]
\centering
\caption{Experimental design for the first cycle. Successive cycle overlap is indicated by ellipses. A, B, X, and Y are biology topics}\label{tab:design}
\begin{tabular}{c@{\hskip 1em}c@{\hskip 1em}l@{\hskip 1em}l@{\hskip 1em}c@{\hskip 1em}c@{\hskip 1em}}
\hline
Week & Class & Group 1 & Group 2 & Immediate Tests & Delayed Test \\ \hline
0    & $A X$  &         &         &                &              \\
1    & $B Y$  & $A_{ITS}$   & $A_{Human}$ & $A X$           &              \\
2    & ...   & $B_{Human}$ & $B_{ITS}$   & $B Y$           &              \\
3    & ...   & ...     & ...     & ...            & $A B X Y$   \\ \hline
\end{tabular}
\end{table}

Topics covered in the study were: active transport, biochemical catalysts, carbohydrate function, diffusion, enzyme reactions, facilitated diffusion, interphase, lipid structure, mitosis, osmosis, protein function, and testing biomolecules.
These topics fall under a single state standard, Cells, and there are interrelationships between some topics.
For example, interphase and mitosis are both part of the cell cycle, and facilitated diffusion and osmosis are both kinds of passive transport.
While tutored conditions covered the same topics, there are potential carryover effects between tutoring and classroom conditions, i.e. tutoring may increase classroom test scores on a related topic.

Analysis options for this design include repeated measures ANOVA on test scores and logistic mixed modeling on item correctness.
Our analysis of the first 3 weeks of the study used repeated measures ANOVA \blindit{\cite{Olney2012a}}.
However, the design and setting of the study in school challenge are a challenge for ANOVA, particularly since students increasingly missed sessions as the study progressed.
Additionally,  test items had varying difficulties, and test items were randomly assigned to tests for each participant
Logistic mixed models address these challenges by easily handling missing data and modeling the variability of test item correctness clustered by participant (cf. ability) and clustered by item (cf. difficulty).
The contrast of interest for the effectiveness of ITS relative to Human and Class conditions is therefore the condition by test interaction using correctness of response on each test item as the dependent variable.


%

\subsection{Knowledge Assessments}

The knowledge assessments were multiple-choice tests where items were either obtained from previous state standardized tests across the U.S. or were created by a researcher for each topic. 
The ratio of researcher-created to standardized items was approximately 2:1.
The researcher who prepared the knowledge tests had access to the topics, the list of concepts for each topic, the biology textbook, and standardized test items. Content from the lectures, scaffolding moves, and other aspects of the ITS condition were not made available to the researcher. 
The researcher was also blind to condition, meaning that the researcher did not know what topics or items students were subsequently tutored on.

Twelve item pre- and post-tests were administered at the beginning and end of each tutoring session for both the ITS and Human conditions to assess prior knowledge and immediate learning gains, respectively. 
Half of the items on each test were on a tutored topic and the other half on an untutored topic (Class condition) as described in Section~\ref{sec:design}.
Test items were randomized across pre- and post-tests, and the order of presentation for individual questions was randomized across students.

Students also completed a 48-item delayed post-test on the third week of each cycle.
Half of the items on this test were previously seen by students (e.g., on the immediate pre- or post-test) and half the items were new but on the same topics. Order of presentation of individual items was randomized across students.

\subsection{Procedure}
Students and parents provided consent prior to the start of the experiment. Students were tested and tutored in groups of two to four in a spare classroom.
The procedure for each tutorial session involved (a) students completing the pretest for 10 minutes (b) a tutorial session with either the ITS or the human tutor for 35 minutes, and (c) the immediate post-test for 10 minutes (all times approximate). The delayed posttest occurred one week after all tutoring was complete.

Interactions with each human tutor occurred in groups, which does not appear to reduce the effectiveness of typical tutors \cite{Ma2014}. 
To obtain a degree of variability in tutoring styles, four human tutors participated in the study on different shifts. The human tutors were provided with the topic to be tutored, the list of ITS concepts for the topic, and the biology textbook students were using in class. Each tutor was an undergraduate major or recent graduate in biology. Before the study, each tutor participated in a one-day training session provided by a nonprofit agency that trains volunteer tutors for local schools.

Students 
interacted with the ITS one-on-one using researcher-provided laptops.
Students wore headphones which prevented them from being distracted by other students. 
A researcher was present in the room to ensure students stayed on task and to start the ITS for each group of students.

\section{Results}

No students were excluded from analysis, and no outliers were removed or transformed.
One student participated in only the human tutor (Human) and classroom-only (Class) conditions, and another student participated in only the ITS and Class conditions, i.e. both students participated in only one session of the study.
These students were retained to prevent bias in the results.
All other students participated in all conditions but not all sessions; generally 75-90\% of students attended sessions in a given week, as shown in Table~\ref{tab:n}.

\begin{table}[htb]
\noindent\begin{minipage}[t]{0.4\textwidth}%
\raggedright
\caption{N per week\hspace*{1.8cm}}\label{tab:n}
\begin{tabular}{@{\extracolsep{4pt}}cccc@{}}
\hline
Cycle & Week 1 & Week 2 & Week 3 \\ \hline
1     & 33     & 31     & 30     \\
2     & 25     & 29     & 26     \\
3     & 25     & 28     & 26     \\ \hline
\end{tabular}
\vspace{1ex}

{\raggedright \textit{Note:} $N=34$ \par}
\end{minipage}%
\begin{minipage}[t]{0.6\textwidth}%
\caption{Assessment descriptive statistics.\hspace*{1.2cm}}\label{tab:desc}
\begin{tabular}{@{\extracolsep{4pt}}lrrrrrr@{}}
\hline
Condition         & \multicolumn{2}{c}{Class}                      & \multicolumn{2}{c}{Human}                      & \multicolumn{2}{c}{ITS}                        \\ \cline{2-3} \cline{4-5} \cline{6-7} 
                  & \multicolumn{1}{c}{M} & \multicolumn{1}{c}{SD} & \multicolumn{1}{c}{M} & \multicolumn{1}{c}{SD} & \multicolumn{1}{c}{M} & \multicolumn{1}{c}{SD} \\ \hline
Sessions  & 2.79                  & .59                    & 2.68                  & .73                    & 2.35                  & .85                    \\
Researcher items  & .71                   & .09                    & .78                   & .14                    & .63                   & .18                    \\
Standardized items & .31                   & .04                    & .24                   & .14                    & .40                   & .16                    \\
Pre-test          & .44                   & .50                    & .43                   & .50                    & .41                   & .49                    \\
Post-test         & .49                   & .50                    & .63                   & .48                    & .64                   & .48                    \\
Delayed test      & .40                   & .49                    & .56                   & .50                    & .55                   & .50                    \\ \hline
\end{tabular}
\end{minipage}%
\end{table}

Table~\ref{tab:desc} shows assessment variable means and standard deviations.
Each student could participate in tutoring conditions once per cycle or 3 times total. 
Most students experienced each condition 2-3 times, with the ITS having the lowest mean number of sessions.
In terms of test items, which were randomized, 
researcher-created and standardized-item proportions were approximately 7:3 overall, with the ITS having the highest proportion of standardized items.

We fit a logistic mixed model and conducted statistical tests at $\alpha=.05$ to answer our first two research questions.
The model with fixed effects was $correctness \sim condition * test$, where condition was Class, Human, or ITS, and test was Pre-test, Post-test, or Delayed.
The model random effects necessarily include participant as there are multiple test responses per participant, and we further included test item as a random effect.
Full specification of the random effects followed recommendations to use the maximal random effects that resulted in model convergence \cite{Barr2013}.
The fixed effects were entered as slopes for the random effects, $(condition * test|participant) + (condition * test|item)$, and slope terms were removed until the model converged, first removing the correlation between slope and intercept and then removing the highest order term.
The decision to remove terms of the same order was made based on diagnostic reports returned by the R package \texttt{glmmTMB} \cite{Brooks2017}, i.e. the term with the most diagnostic red flags was removed first.
The converging model was an intercepts-only model, $correctness \sim condition * test + (1|participant) + (1|item)$.
Comparison of the maximal model and converging model revealed that they had the same significant fixed effects, suggesting that these effects are robust because they hold across nontrivial changes to the random effect structure.

A Type III ANOVA analysis of the converging model \cite{Singmann2023} revealed a significant main effect of condition, $\chi^2(2)=10.83, p=.004$, a significant main effect of test,  $\chi^2(2)=117.13, p<.001$, and a significant interaction between condition and test, $\chi^2(4)=43.50, p<.001$.
To answer our first two research questions, we conducted contrasts within the interaction using Tukey's p-value adjustment \cite{Lenth2025}.
Logistic models intrinsically provide an effect size through odds ratio (OR), or the times more likely an event will occur, which we converted into Cohen's $d$ for comparison to aforementioned effect sizes \cite{BenShachar2020}.

\subsubsection{Contrasts between Conditions.} 
Contrasts of pre-test scores between conditions revealed that there were no significant differences between ITS and Human, $z=.12, p=.992$, between ITS and Class, $z=-.57, p=.836$, or between Human and Class, $z=-.57, p=.836$.
Contrasts of post-test scores between conditions revealed no significant difference between ITS and Human, $z=.62, p=.809$, but revealed significant differences between ITS and Class, $z=4.24, p<.001$, and between Human and Class, $z=3.84, p<.001$, such that ITS was more likely to answer questions correctly on the post-test than Class, $OR=2.24, d=.50$, and Human was more likely to answer questions correctly on the post-test than Class, $OR=2.04, d=.44$.
Contrasts of delayed test scores between conditions revealed no significant difference between ITS and Human, $z=-.34, p=.939$, but revealed significant differences between ITS and Class, $z=4.24, p<.001$, and between Human and Class, $z=4.42, p<.001$, such that ITS was more likely to answer questions correctly on the delayed test than Class, $OR=2.04, d=.45$, and Human was more likely to answer questions correctly on the delayed test than Class, $OR=2.11, d=.47$.

\subsubsection{Contrasts within Conditions.} 
Contrasts of test scores within conditions mirrored the pattern of results between conditions.
In the ITS condition, there were significant differences between pre-test and post-test, $z=7.60, p < .001$, pre-test and delayed test, $z=3.90, p < .001$, and between post-test and delayed test,  $z=-3.88, p < .001$, such that the ITS condition was more likely to correctly answer questions on the post-test than the pre-test, $OR=3.13, d=.71$, and more likely to correctly answer questions on the delayed test than the pre-test, $OR=1.77, d=.36$, but less likely to correctly answer questions on the delayed test than the post-test, $OR=.56, d=-.35$.
In the Human condition, there were significant differences between pre-test and post-test, $z=7.71, p < .001$, pre-test and delayed test, $z=4.34, p < .001$, and post-test and delayed test,  $z=-3.12, p = .005$, such that the Human condition was more likely to correctly answer questions on the post-test than the pre-test, $OR=2.90, d=.66$, and more likely to correctly answer questions on the delayed test than the pre-test, $OR=1.86, d=.39$, but less likely to correctly answer questions on the delayed test than the post-test, $OR=.63, d=-.27$.
In the Class condition, there were no significant differences between pre-test and post-test, $z=2.27, p = .06$, or between the pre-test and delayed test, $z=-1.86, p = .150$. 
However, there was a significant difference between the delayed test and the post-test,  $z=-3.57, p = .001$, such that the Class condition was less likely to correctly answer questions on the delayed test than the post-test, $OR=.62, d=-.30$.

\subsubsection{Exploratory Analysis.} 
An additional exploratory analysis was performed to investigate whether the variable delay between tutoring sessions and the delayed test affected test scores, i.e. whether items on topics covered more recently were more likely to be answered correctly. 
The model was refit with a nominal delay term (Week 1, Week 2) as a fixed effect.
The delay term was not significant, $\chi^2(1)=.94, p=.333$, and the pattern of significant fixed effects was unchanged.

\subsection{Discussion}

Our research questions in this study were (1) is learning in the ITS condition different from the classroom control and (2) is learning in the ITS condition different from subject matter expert tutors?
Results of the study clearly answer these two questions.
For the first question, the ITS condition improved 
learning on both the immediate post- and delayed-test, but the Class showed no significant improvement in learning. 
These results suggest that the ITS condition is more effective than the Class condition at promoting learning.
For the second question, 
both ITS and Human conditions had approximately the same effects on learning.
The ITS had a slightly larger effect on pre- to post-test ($d=.71$) compared to Human ($d=.66$), and Human had a slightly larger effect on pre- to delayed test ($d=.39$) compared to ITS ($d=.36$), but the ITS and Human conditions were not significantly different at post- or delayed tests.
Additionally, the pre- to post-test effect for the ITS was in the 99th percentile for ITS that use both standardized and researcher-created test items \cite{Kulik2016}, and the pre- to post-test effect for the human tutors is in the 99th percentile of meta-analytic effects \cite{Hartley1977,Cohen1982,Mathes1994,Ritter2009,VanLehn2011} when the 2-sigma studies confounding mastery learning with tutoring are removed.
These results suggest that the ITS condition and the Human condition were both equally effective at promoting learning and highly effective at promoting learning.

While these results are very positive, the lack of contrast between Human and ITS conditions is also perhaps the greatest limitation of the study.
Our intuition for this research is that expertise in tutoring makes a contribution to tutoring effectiveness beyond subject matter expertise.
However, we were not able to directly test this intuition in the study, and our results are consistent with the claim that expertise in the domain is more important than tutoring expertise \cite{VanLehn2011}.
If we had included another tutoring system, matched for content, that modeled the behaviors of novice human tutors, then a contrast with the ITS condition could have further informed our understanding of the role of tutoring expertise.
Likewise, if we had included expert human tutors with the experience criteria described in Section~\ref{sec:intro}, that condition would have provided another useful contrast.
However, we were limited by the constraints of the school we were working in (i.e. we could only pull students out of certain classes) as well as the resources available to us (i.e. creating a novice ITS condition matched for content was not part of our project).

Our results have additional limitations.
The Class condition was based on a single teacher, the only Biology I teacher in the school.
A better (or worse) teacher would have affected pre-test scores and corresponding comparisons within and between conditions.
Additionally, though we used a mixture of standardized and researcher-created test items, we were unable to analyze test performance on them separately, because these item types were both unevenly distributed across topics and randomly assigned across tests (i.e. a topic may have only researcher-created items or a student may have all standardized items on the pre-test).
Our results capture variations in difficulty by using test item as a random effect, but separate analyses by test type would allow a tighter comparison with reported meta-analytic effects for both ITS and human tutoring.

One of the greatest challenges in expert tutoring research is the construction of assessments.
The expert tutors we have studied do not use curriculum scripts, i.e. pre-planned teaching agendas, but rather base instruction on student needs dynamically \blindit{\cite{Olney2010b}}. 
It is impossible to prepare learning assessments in advance for dynamic instruction, and the standard practice is to control for content and use the same assessments across conditions.
We argue that controlling for content across conditions effectively forces the use of a curriculum script and would create a handicap for expert human tutor effectiveness.
Recent advances in generative AI for assessment suggest that large language models can dynamically produce multiple choice questions and that these questions have comparable quality and psychometric properties to human-authored questions on the same topic \cite{Olney2023a,Bhandari2024}.
If effective learning assessments can be dynamically generated by AI, then future studies could study both human tutors who were experts in the domain but not in tutoring and expert human tutors in their natural contexts, across topics.
Such research would better inform our understanding of the role of expertise in tutoring: not just what expert human tutors do, but how those differences translate into effectiveness. 



\begin{credits}
\subsubsection{\ackname} This research was supported by \blindit{the National Science Foundation (NSF) (HCC 0834847, DRL 1108845, DUE 1918751) and Institute of Education Sciences (IES), U.S. Department of Education (DoE), through Grant R305A080594. Any opinions, findings and conclusions, or recommendations expressed in this paper are those of the authors and do not necessarily reflect the views of NSF, IES, or DoE}.

\subsubsection{\discintname}
The authors have no competing interests to declare that are
relevant to the content of this article.
\end{credits}
%
%
%
\bibliographystyle{splncs04}
\bibliography{references}
\end{document}